\documentclass[conference]{IEEEtran}
\IEEEoverridecommandlockouts
\usepackage[bottom=1.02in,top=0.71in,left=0.64in,right=0.64in]{geometry}
\usepackage{amsmath}
\usepackage{amsfonts}
\usepackage{amssymb}
\usepackage{algorithmic}
\usepackage[ruled,vlined]{algorithm2e}
\usepackage{caption}
\usepackage{mathtools}  
\usepackage{amsthm,bm}
\usepackage[]{algorithm2e}
\usepackage{booktabs}
\usepackage{accents}
\usepackage{subfigure}
\usepackage{tabulary}
\usepackage{footnote}
\usepackage[export]{adjustbox}
\usepackage{booktabs}
\usepackage[justification=centering]{caption}
\usepackage{comment}
\newtheorem{remark}{Remark}
\usepackage{cases}
\usepackage{graphicx}
\usepackage{cite}
\usepackage{multicol}
\usepackage{xcolor}
\usepackage{graphicx}
\graphicspath{{./}{figures/}}
\newcommand\numeq[1]%
  {\stackrel{\scriptscriptstyle(\mkern-1.5mu#1\mkern-1.5mu)}{=}}
\usepackage{stfloats}
\usepackage{cuted}
\usepackage{lipsum}  

\usepackage{dsfont}
\usepackage{amsmath,amssymb}

\usepackage{optidef}
\usepackage{amsmath}

\title{Codebook-based Uplink Transmission Enhancement in 5G Advanced: Sub-band Precoding}

\author{\IEEEauthorblockN{Liu Cao$^*$$^\dagger$, Yahia Shabara$^\dagger$, Parisa Cheraghi$^\dagger$}
\IEEEauthorblockA{{$^*$Department of Electrical and Computer Engineering, University of Washington, Seattle, WA, USA}\\ {$^\dagger$MediaTek USA Inc., San Diego, CA, USA}\\
Emails: liucao@uw.edu, \{yahia.shabara, parisa.cheraghi\}@mediatek.com}

\thanks{This work was completed by L. Cao during his internship.}

}

\begin{document}

\maketitle
\thispagestyle{empty}
\begin{abstract}
The transformative enhancements of fifth-generation (5G) mobile devices bring about new challenges to achieve better uplink (UL) performance. Particularly, in codebook-based transmission, the wide-band (WB) precoding and the legacy UL codebook may become main bottlenecks for higher efficient data transmission. In this paper, we investigate the codebook-based UL single-layer transmission performance using fully coherent antenna ports in the context of sub-band (SB) precoding. We analyze the SB precoder selection criteria and design an UL codebook used for SB precoding by increasing the number of relative phase shifts of each port. Via link-level simulations, we verify that the UL SB precoding can improve up to 2 dB performance gain in terms of the block error rate (BLER) compared with the UL WB precoding which is the current UL precoding scheme. We also show that UL performance gain is sensitive to the SB size selection as well as the relative phase shift diversity.

\end{abstract}


\section{Introduction}
\label{sec:intro}
 While 5G downlink (DL) often receives a lot of attention due to the desire for faster download speeds, the uplink (UL) is equally crucial, especially with the surge in content creation, cloud computing, and IoT devices that need to transmit data back to the network. Currently, UL transmission in 5G is achieved through two schemes \cite{5gBullet}: Non-codebook-based transmission or codebook-based transmission. In non-codebook-based transmission, the user equipment (UE) measures DL channel state information (CSI) reference signal (CSI-RS) to generate its own precoding weights for the physical uplink shared channel (PUSCH). These precoding weights are not constrained to a codebook standardized by the 3rd Generation Partnership Project (3GPP). By contrast, in codebook-based transmission, the UE transmits the PUSCH using precoding weights that have been selected from a codebook standardized by 3GPP. These standardized codebooks are carefully designed to satisfy the specific antenna configurations.





In particular, Release 16/17 allows for UL transmission using 1,2, or 4 antenna ports (TX) with specified codebooks \cite{3gpp.38.211}. However, restricting the number of UL TX ports to only up to 4 limits the achievable performance, which causes an imbalance between the UL and DL capacities. Meanwhile, increasing the number of TX ports at the UE also enables higher diversity gain. Thus, Release 18 supports 8TX UL transmission to cope with the above issues \cite{jin2023massive, R1-2204692}. However, the current codebook-based UL transmission is via wide-band (WB) precoding, which may significantly limit the channel capacity of the whole band, especially under high bandwidth. Instead, sub-band (SB) precoding might be an option to improve the performance gain \cite{R1-2205221}. Meanwhile, most existing research work \cite{dong2022distributed, pang2019uplink,xie2022uplink, wang2023uplink,yin2022routing,yin2020multiplexing} investigates the UL transmission performance in the context of WB precoding. Hence, it is necessary to study the impact of the SB precoding on UL performance using different numbers of TX. On the other hand, the legacy UL codebooks include a limited number of precoding matrices for selection, which may limit the SB precoding performance gain if the legacy UL codebook is still adopted. Thus designing a new codebook used for SB precoding is also in demand. 
 
Motivated by the aforementioned issues, this paper investigates the codebook-based UL single-layer transmission performance using fully coherent antenna ports in the context of SB precoding. We analyze the SB precoder selection criteria and design an UL codebook used for SB precoding by increasing the number of relative phase shifts of each port. The major contributions of this paper are summarized as follows: 
\begin{itemize}
\item We investigate the single-layer performance for different numbers of TX. Particularly, the 8TX performance is also tested based on the latest UL codebook that is standardized by 3GPP Release 18. 
\item We verify that SB precoding improves the performance gain compared with WB precoding which is the current UL precoding scheme. Meanwhile, we illustrate that the performance gain is sensitive to the SB size selection as well as the relative phase shift diversity.

\end{itemize}

The rest of this paper is organized as follows: Section \ref{sec:sys_arc}  introduces the procedure for codebook-based UL transmission and recaps the UL codebook types. The codebook design for the single-layer transmission is illustrated in Section \ref{sec:sys_model}. In Section \ref{sec:sim}, we present simulation results with different TX cases. Finally, Section \ref{sec:con} draws conclusions for this paper. 




\section{System Architecture}
\label{sec:sys_arc}
\subsection{Codebook-based UL transmission }
\begin{figure}[ht]
    \centering
\includegraphics[width=.48\textwidth]{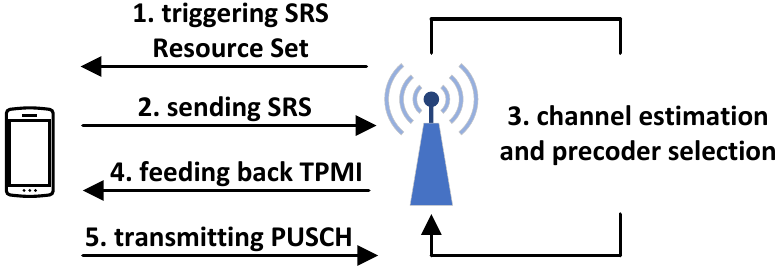}
    \caption{Procedure for codebook-based UL transmission.}
    \label{fig:E2Esysarch}
\end{figure}

Fig. \ref{fig:E2Esysarch} illustrates the procedure for codebook-based UL transmission. The gNodeB first uses the physical downlink control channel (PDCCH) to trigger the sounding reference signal (SRS) resource set of each antenna panel\footnote{The maximum number of SRS resources equates to the number of antenna panels, e.g., the single SRS resource set includes 1 SRS resource if the UE has 1 antenna panel and 2 SRS resources if the UE has 2 antenna panels.}. The UE transmits the SRS from each of its antenna ports once the SRS resource set has been triggered. The gNodeB then compares the SRS transmissions belonging to each SRS resource to select one of the antenna panels\footnote{The end-user may be shielding one panel with a hand wrapped around the UE. The gNodeB thereby should detect the weaker signal and instruct the UE to use the other panel.}. Afterward, the gNodeB evaluates the SRS transmissions to determine an appropriate rank (number of layers) and precoding matrix for the selected antenna panel,  which both depend upon the propagation channel between the UE and the gNodeB. The gNodeB attempts to identify a precoding matrix that will maximize the rank and the received signal-to-noise ratio (SNR). 

The gNodeB then uses the DL control information (DCI) Format $0\_1$ to allocate the PUSCH resources. The DCI includes an SRS resource indicator (SRI) that instructs the UE to use a specific antenna panel, meanwhile it also contains a field that instructs the UE to use a specific combination of the precoding matrix and the number of layers. The precoding matrix is identified using its transmitted precoding matrix indicator (TPMI). Finally, the UE is able to use the resource allocation received on the PDCCH to transmit the PUSCH while using the selected antenna panel, the number of layers and the precoding matrix.

\subsection{UL Codebook types: recap}
Codebooks are usually presented as a list of matrices where each matrix defines a set of precoding weights. The dimensions of each matrix depend on the number of layers and the number of antenna ports. The number of rows of the precoding matrix is equal to the number of antenna ports, while the number of columns is equal to the number of layers. This allows matrix multiplication between the precoding matrix and a column vector which includes the input symbols from each layer. Multiple codebooks are defined to satisfy the range of layers and the range of antenna ports. Each precoding matrix is also preceded by an amplitude scaling factor which is used to manage the total output power according to the number of layers and number of antenna ports\footnote{For instance, a precoding matrix that includes 4 elements may be preceded by a scaling factor of $\frac{1}{2}$. This indicates that the power of each entry is $\frac{1}{4}$ and thus the total power keeps unity.}. 

When a device has multiple antenna ports, those ports can be fully coherent, non-coherent, or partially coherent \cite{5gBullet}. The UE is able to control the relative phase of signals transmitted by fully coherent antenna ports, while it is not able to control the relative phase of signals transmitted by non-coherent antenna ports. The partially coherent ports include fully coherent ports and non-coherent ports. In this paper, we focus on the fully coherent port case, where the first antenna port transmits the single-layer directly, while the other antenna port(s) transmits the single-layer with different phase shifts.

\section{codebook design for the single-layer}
\label{sec:sys_model}

\subsection{SB precoder selection criteria}
We consider a typical UL MIMO system where the UE is equipped with $N_t$ antennas, and the gNodeB uses $N_r$ antennas. The wide-band channel is divided into a sub-band set $L$, and each sub-band includes a sub-carrier set $C$ that has the same number of sub-carriers. The $N_r \times 1$ received signal at the $c$th sub-carrier in the $l$th sub-band $\mathbf{y}_{l,c}$ is given by
\begin{equation}\begin{aligned}\label{eq:y_n1}
\mathbf{y}_{l,c} =  \mathbf{H}_{l,c}\mathbf{W}_l^{(CB)}s_{l,c} + \mathbf{n}_{l,c},
\end{aligned}\end{equation}
where $s_{l,c} \in \mathbb{C}$, $\mathbf{H}_{l,c} \in \mathbb{C}^{N_r \times N_t}$ and $\mathbf{n}_{l,c} \in \mathbb{C}^{N_r \times 1}$ are the signal, the channel matrix and the addictive Gaussian noise at the $c$th sub-carrier in the $l$th sub-band, respectively.     
$\mathbf{W}_l^{(CB)} \in \mathbb{C}^{N_t \times 1}$ is the selected precoding matrix for the $l$th sub-band from a given codebook $\mathbf{W}^{(CB)}$. The selection of $\mathbf{W}_l^{(CB)}$ is determined by the channel covariance matrix of the $l$th sub-band, which can be represented by the average of channel covariance matrices of all sub-carriers in the $l$th sub-band:
\begin{equation}\begin{aligned}\label{eq:c_l}
\mathbf{C}_{l} = \mathbf{H}_{l}^{H}\mathbf{H}_{l} = \frac{1}{|C|}\sum_c\mathbf{C}_{l, c} = \frac{1}{|C|}\sum_c\mathbf{H}_{l,c}^{H}\mathbf{H}_{l,c},
\end{aligned}\end{equation}
where $\mathbf{C}_{l} \in \mathbb{C}^{N_t \times N_t}$, $C$ represents the number of sub-carriers, and $\mathbf{H}_{l,c}^{H}$ is the conjugate transpose of $\mathbf{H}_{l,c}$. Through the singular value decomposition (SVD), 
$\mathbf{C}_{l}$ can be decomposed into $\mathbf{V}_{l}\Sigma_l\mathbf{V}_{l}^H$, where $\Sigma_l \in \mathbb{R}^{+}$ (rank = 1 due to single-layer) and $\mathbf{V}_{l} \in \mathbb{C}^{N_t \times 1}$. As a result, the $N_r \times 1$ received signal in the $l$th sub-band can be expressed as
\begin{equation}\begin{aligned}\label{eq:y_n2}
\mathbf{y}_l =  \mathbf{U}_l\sqrt{\Sigma_l}\mathbf{V}_l^{H}\mathbf{W}_l^{(CB)}s_l + \mathbf{n}_l,
\end{aligned}\end{equation}
 where $s_{l} \in \mathbb{C}$ and $\mathbf{n}_{l} \in \mathbb{C}^{N_r \times 1}$ are the signal and addictive Gaussian noise in the $l$th sub-band, respectively. $\mathbf{U}_l \in \mathbb{C}^{N_r \times 1}$ is an unitary matrix. By multiplying $\mathbf{U}_l^H$, the received signal in the $l$th sub-band is
\begin{equation}\begin{aligned}\label{eq:z_n}
z_l =  \mathbf{U}_l^H \mathbf{y}_l = \sqrt{\Sigma_l}\mathbf{V}_l^{H}\mathbf{W}_l^{(CB)}s_l + \mathbf{U}_l^H\mathbf{n}_l.
\end{aligned}\end{equation}
As a result, the post-processing SNR in the $l$th sub-band is given by
\begin{equation}\begin{aligned}\label{eq:snr_n}
\Gamma_l &=  \frac{P_l|\sqrt{\Sigma_l}\mathbf{V}_l^{H}\mathbf{W}_l^{(CB)}|^2}{|\mathbf{U}_l^H\mathbf{n}_l|^2},
\end{aligned}\end{equation}
where $|*|$ is the magnitude of a complex number, and $P_l$ is the transmit power in the $l$th sub-band. 
\begin{remark}
 As $\mathbf{W}_l^{(SVD)} = \mathbf{V}_l$, then $|\sqrt{\Sigma_l}\mathbf{V}_l^{H}\mathbf{W}_l^{(SVD)}|^2=\Sigma_l$. While $|\sqrt{\Sigma_l}\mathbf{V}_l^{H}\mathbf{W}_l^{(CB)}|^2 = \Sigma_l|\mathbf{V}_l^{H}\mathbf{W}_l^{(CB)}|^2 < \Sigma_l$ as $||\mathbf{V}_l^{H}||_F = ||\mathbf{W}_l^{(CB)}||_F = 1$, where $||*||_F$ is the Frobenius norm of a matrix. Thus $\mathbf{W}_l^{(SVD)}$ is the best SB precoder solution that any $\mathbf{W}_l^{(CB)}$ cannot outperform.
\end{remark}
Each time gNodeB performs a full search of available precoding matrices from the given codebook and selects one that maximizes $|\sqrt{\Sigma_l}\mathbf{V}_l^{H}\mathbf{W}_l^{(CB)}|$. Afterward, gNodeB sends the UE a TPMI list corresponding to all sub-bands. 
\begin{remark}
The WB precoding is a special case of the SB precoding, where $\mathbf{W}_l^{(CB)}$ keeps the same for each $l$, i.e., $\mathbf{W}_l^{(CB)} = \mathbf{W}_1^{(CB)}$ for all $l$.  Then  $|\mathbf{V}_l^{H}\mathbf{W}_l^{(CB)}|^2 \geq |\mathbf{V}_l^{H}\mathbf{W}_1^{(CB)}|^2$ for all $l$ must be satisfied. Thus, $\Gamma_l$ using the SB precoding must be higher than that using the WB precoding.
\end{remark}






\subsection{UL Codebook design for fully coherent ports}


The codebooks for 2TX and 4TX using fully coherent ports 
have been specified in Release 16/17 \cite{3gpp.38.211}. where 4 and 16 specific precoding matrices for 2TX and 4TX single-layer are shown, respectively. However, the codebook for 8TX using fully coherent ports is different from that for 2/4TX because it adopts the DL Type I codebooks in Release\cite{3gpp.38.214}. Type I codebooks are 2D DFT beams with oversampling enabled to provide some additional spatial granularity. The UL 8TX beam vector is expressed as
\begin{equation}\begin{aligned}\label{eq:v_i}
\bm{v_i} =  \bm{v}^{(H)}_{i_{h}} \otimes \bm{v}^{(V)}_{i_{v}},
\end{aligned}\end{equation}
where $\otimes$ is Kronecker product. The horizontal beam
\begin{equation}\begin{aligned}\label{eq:v_ih}
\bm{v}^{(H)}_{i_{h}} =   \Bigl(1 \quad e^{j\frac{2\pi i_h}{O_1N_1}} \quad e^{j\frac{2\pi i_h 2}{O_1N_1}} \quad  \cdots \quad e^{j\frac{2\pi i_h(N_1-1)}{O_1N_1}}\Bigl)^T,
\end{aligned}\end{equation} 
where $N_1$ and $O_1$ are the number of antennas and the number of beams in the horizontal direction, respectively. The vertical beam
\begin{equation}\begin{aligned}\label{eq:v_iv}
\bm{v}^{(V)}_{i_{v}} =   \Bigl(1 \quad e^{j\frac{2\pi i_v}{O_2N_2}} \quad e^{j\frac{2\pi i_v 2}{O_2N_2}} \quad  \cdots \quad e^{j\frac{2\pi i_v(N_2-1)}{O_2N_2}}\Bigl)^T,
\end{aligned}\end{equation}
where $N_2$ and $O_2$ are the number of antennas and the number of beams in the vertical direction, respectively. And $\bm{i}_h \in \{0, 1, 2, \cdots, O_1N_1-1\}$ while $\bm{i}_v \in \{0, 1, 2, \cdots, O_2N_2-1\}$.
The precoder for 8TX single-layer using fully coherent ports is given by 
\begin{equation}\begin{aligned}\label{eq:W}
W_{i_h, i_v, n} =   \frac{1}{2\sqrt{2}} \Bigl(\bm{v_i}^T  \quad \varphi_n \bm{v_i}^T \Bigl)^T,
\end{aligned}\end{equation}
where $\varphi_n = e^{\frac{j\pi n}{2}}$, and $n = \{0, 1, 2, 3\}$. For UL 8TX, it has been agreed that there is no need to support oversampling factors, i.e., assume that $O_1 = O_2= 1$. As a result,  $|i_h| = N_1$ and $|i_v| = N_2$\footnote{Due to the cross-polar antennas, $N_1N_2 = 4$ in the 8TX case. The possible configurations are $N_1 = 4$ and $N_2 = 1$ or $N_1 = 2$ and $N_2 = 2$.}. We can thereby justify that the total number of available precoding matrices for the 8TX single-layer using fully coherent ports is $|i_h||i_v||n| = 4N_1N_2=16$.

Note that the legacy codebook above is only aimed at the UL wide-band precoding, whose precoding matrices are limited for selection. The sub-band precoding performance may be limited by the legacy precoder choices. Thus, as \textbf{Algorithm 1} shows, we design an SB UL codebook for the single-layer, where $2^{M_i}$ represents the number of relative phases for the $i$th port. The proposed codebook increases the precoding diversity by uniformly increasing the relative phase diversity of each port. Thus each sub-band can benefit from increased flexibility with more precoding weights.

\begin{algorithm}[h]
 \KwIn{Number of TX $N_{t}$, and number of precoders $2^{N_W}$.}
 \KwOut{Codebook $\mathbf{W}^{(CB)}$.}

\begin{algorithmic}
\STATE \small{$\mathbf{W}^{(CB)} = \Big\{\frac{1}{\sqrt{N_{t}}}\Big(1 \quad e^{j\frac{2\pi m_2}{2^{M_2}}} \quad \cdots \quad e^{j\frac{2\pi m_{N_{t}}}{2^{M_{N_{t}}}}}\Big)^T,$  where $\sum\limits_{i = 2}^{N_{t}}M_i = N_W \ \text{and} \ m_i \in \{0, 1, \cdots, 2^{M_i}-1\}  \Big\} $.}
  \end{algorithmic}
  \caption{SB UL codebook for the single-layer using fully coherent ports.}
\end{algorithm}

\begin{remark}
  As the relative phase diversity of each port is increased,   $|\mathbf{V}_l^{H}\mathbf{W}_l^{(CB,proposed)}|^2 \geq |\mathbf{V}_l^{H}\mathbf{W}_l^{(CB,legacy)}|^2$ for all $l$ must be satisfied. Thus, $\Gamma_l$ using the proposed codebook must be higher than that using the legacy codebook.  
\end{remark}

\section{Simulation}
\label{sec:sim}

In this section, we test the SB precoding performance for different numbers of TX by conducting various simulations via the Python-based link-level simulator (LLS) developed by MediaTek. The main simulation setup is summarized in Table \ref{tab:sim_para} \footnote{$N_1 = 4$ and $N_2 = 1$ in the 8TX simulations.}. We add the WB precoding using the legacy codebook which is the current codebook-based UL precoding scheme as the baseline to justify the effectiveness of the SB precoding. Meanwhile, we also add the WB precoding using SVD as well as the SB precoding using SVD to provide more insights.

\begin{table}[h]
 \centering
 \caption{Main simulation setup. 
}\label{tab:sim_para}
\resizebox{.48\textwidth}{!}{\begin{tabular}{ |c|c|c|c|c|c|c| } 
\hline
Total Bandwidth (MHz) &  100 & Carrier Frequency (GHz) &  3 \\
\hline
Numerology  & 1 & MCS  & 22 \\
\hline
Number of total RBs  & 270 & $\#$ of TBs/Simulation  & 1500 \\
\hline
Channel Type  & CDL-A  & $\#$ of RX &  8 \\
\hline
Delay Spread (ns)  & 300  & UE Speed (km/h) &  6 \\
\hline
Channel Estimation & MMSE & Channel Coding &  LDPC \\
\hline
\end{tabular}}
\end{table}

\begin{figure}[t]
    \centering
\includegraphics[width=.48\textwidth]{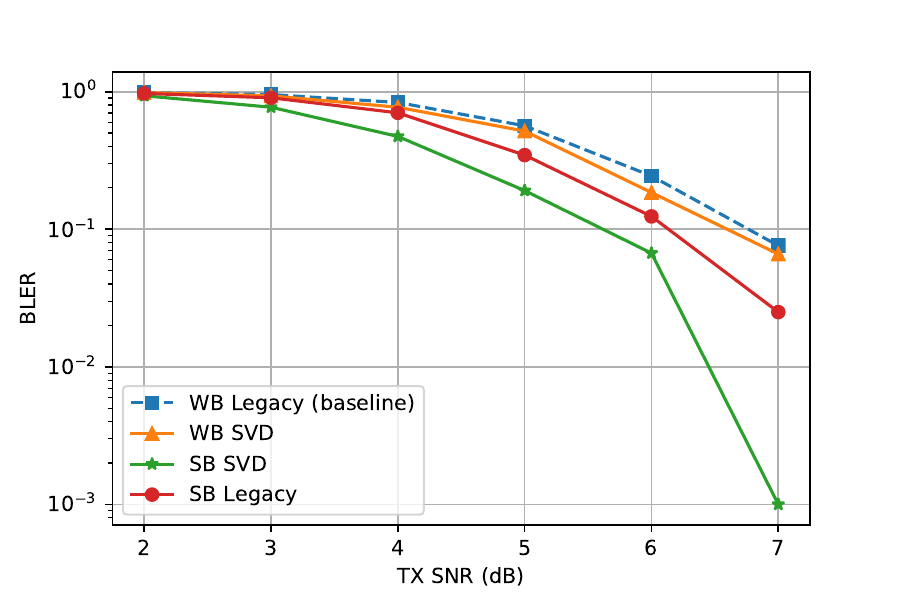}
    \caption{2TX BLER comparison.}
    \label{fig:BLER2TX}
\end{figure}

\subsection{2TX case}
We first investigate the block error rate (BLER) in terms of SNR using different precoding methods, which is shown in Fig. \ref{fig:BLER2TX}. For SB precoding, the SB size (SBS) is 1 RB, thus the total number of SBs is 270. By comparing SB legacy to WB legacy, the SB legacy improves the baseline performance by increasing around 0.5 dB gain, whose reason can be indicated by Remark 2. By comparing SB legacy to WB SVD, the SB precoding matters more than the WB precoding accuracy in the single-layer transmission. Besides, as there is a big BLER gap between WB legacy and SB SVD, we can predict that SB precoding may have significant room for possible performance improvement.

\begin{figure}[t]
    \centering
\includegraphics[width=.48\textwidth]{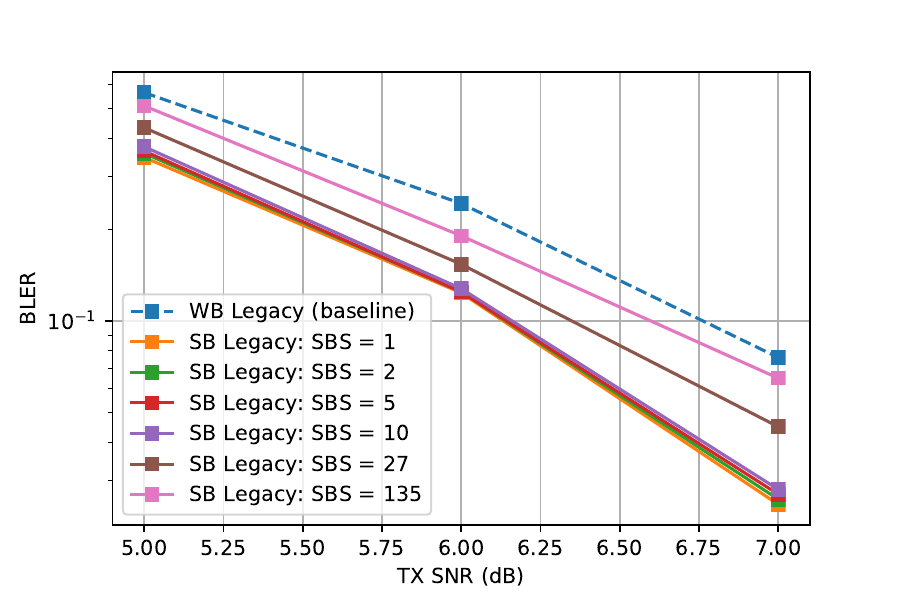}
    \caption{2TX BLER in terms of SBS using SB legacy.}
    \label{fig:BLER2TXSBS}
\end{figure}

\begin{figure}[t]
    \centering
\includegraphics[width=.48\textwidth]{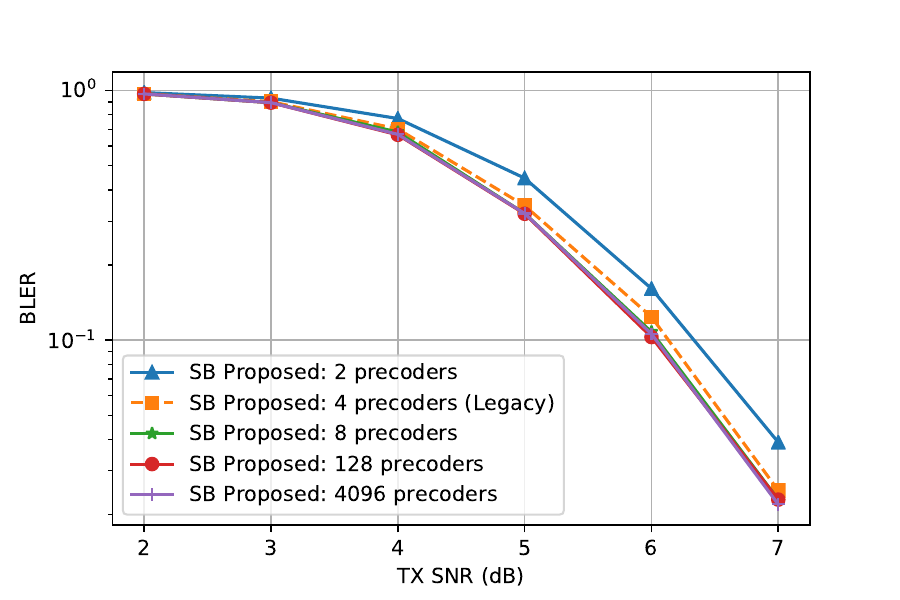}
    \caption{2TX BLER using SB proposed.}
    \label{fig:BLER2TXProposed}
\end{figure}

Afterward, we test the efficiency of SB legacy by using different SBS. As Fig. \ref{fig:BLER2TXSBS} shows, the larger the SBS is, the lower the BLER is. Thus SBS = 1 can achieve the lowest BLER among all SBSs. However, the performances whose SBS ranges from 1 to 10 are quite close.  This is because the spread delay was set to 300 ns in the simulation, the corresponding coherence bandwidth is around 10 RBs. As a result, the selected precoding matrices should be almost the same within every 10 RBs when SBS is less than 10, which increases the unnecessary signaling overhead. Hence SBS = 10 may be a better choice under this simulation setup for higher SB precoding efficiency. 




As is shown in Fig. \ref{fig:BLER2TXProposed}, we further test the efficiency of the proposed SB method by using different phase diversity. Note that the SB legacy is one of SB proposed cases where $2^{N_W} =4$. As Remark 3 indicates, the larger the precoder diversity is, the lower the BLER is. However, the BLER performance is not very sensitive to the larger relative phase shift diversity. This is because it is hard to form a narrower beam by only using 2TX. Therefore, increasing to large phase diversity for 2TX does not help improve the performance much. The $8$-precoder scheme ($N_W = 3$) that increases by around 0.2 dB gain (compared to SB legacy) is good enough in the 2TX case under the given simulation setup. Thus, the proposed 8 precoders can increase by around 0.7 dB if it is compared to the WB legacy scheme.

\subsection{4TX case}
\begin{figure}[t]
    \centering
\includegraphics[width=.48\textwidth]{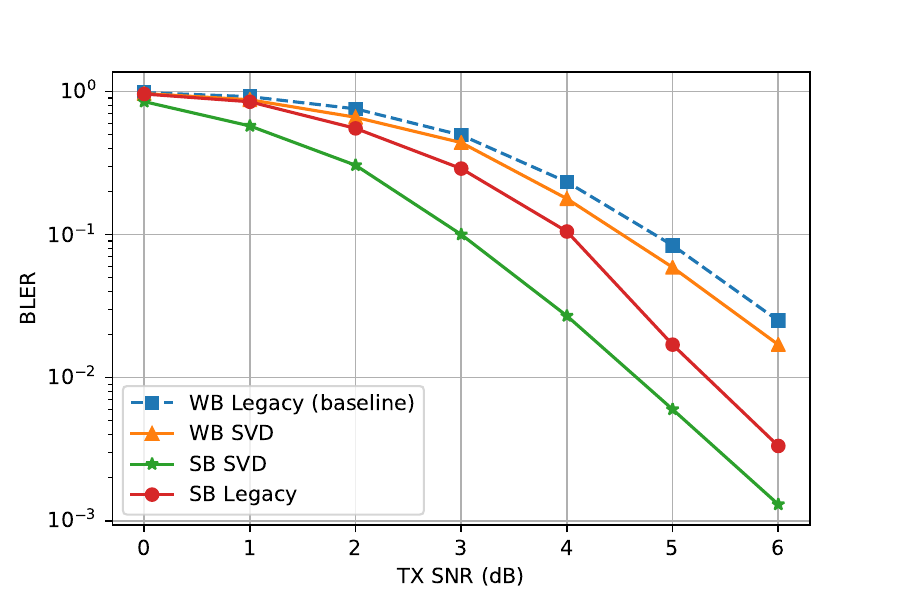}
    \caption{4TX BLER comparison.}
    \label{fig:BLER4TX}
\end{figure}

\begin{figure}[t]
    \centering
\includegraphics[width=.48\textwidth]{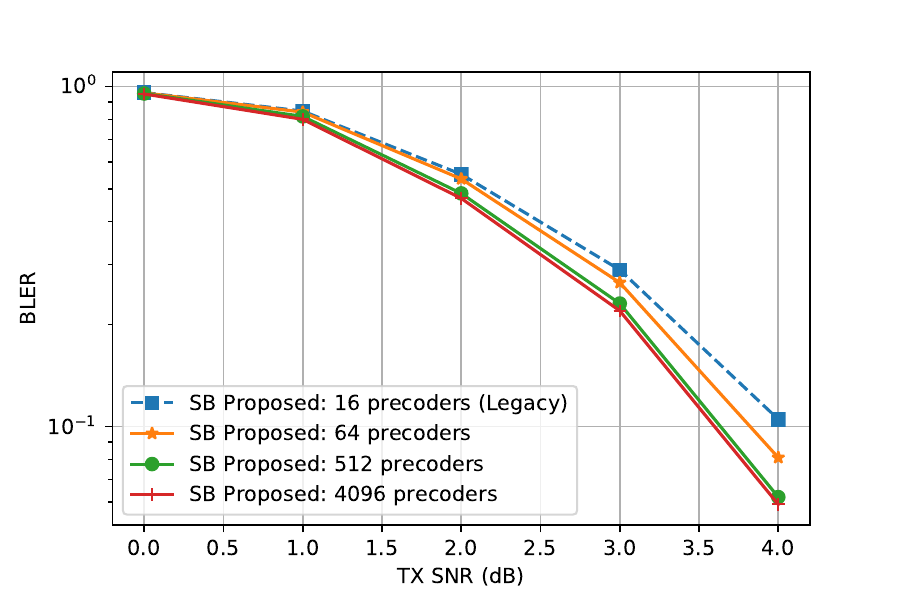}
    \caption{4TX BLER using SB proposed.}
    \label{fig:BLER4TXProposed}
\end{figure}
Similar to the 2TX case, we investigate the 4TX BLER using different precoding methods, which are shown in Fig. \ref{fig:BLER4TX} and \ref{fig:BLER4TXProposed}, respectively. For SB precoding, the SBS is still 1 RB in the 4TX case. The conclusions based on the 4TX simulations are almost the same as those from the 2TX simulations, except for the following difference: 1) By comparing SB legacy to WB legacy, the performance gain with 4TX is higher than that with 2TX. Particularly, 4TX SB legacy can improve the baseline performance by increasing around 1 dB gain in Fig. \ref{fig:BLER4TX}; 2) Increasing to a large phase diversity for 4TX helps improve the performance better than 2TX. The proposed 512 precoders (where $M_2 = M_3 = M_4 = 3$) can increase by around 1.4 dB compared to the WB legacy in Fig. \ref{fig:BLER4TXProposed}.

\subsection{8TX case}

\begin{figure}[t]
    \centering
\includegraphics[width=.48\textwidth]{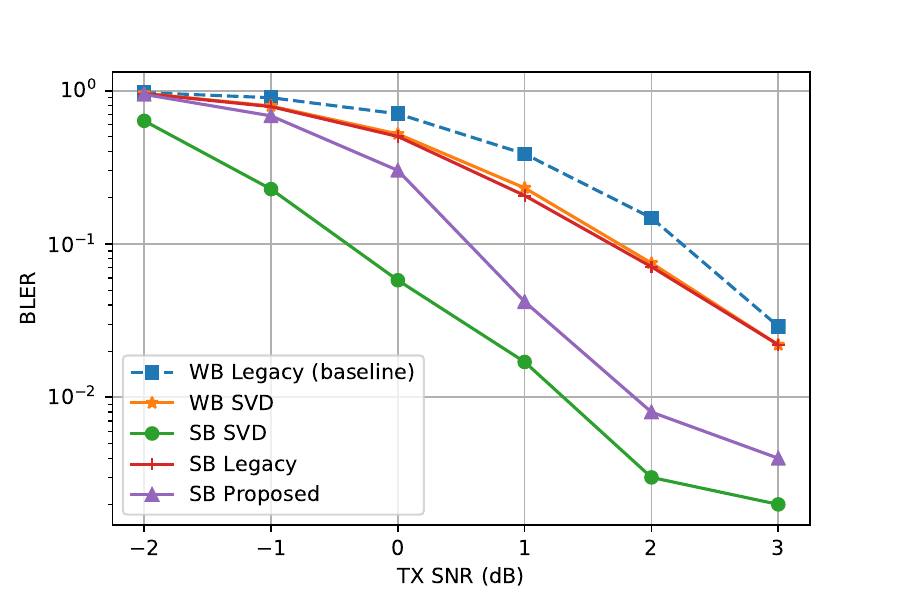}
    \caption{8TX BLER comparison.}
    \label{fig:BLER8TX}
\end{figure}

As Fig. \ref{fig:BLER8TX} shows, we finally investigate the 8TX BLER using different precoding methods. For SB precoding, the SBS is still 1 RB in the 8TX case. In general, the conclusions based on the 2 and 4TX simulations are still applicable for 8TX. However, by comparing SB legacy to WB legacy, it can be found that the performance gain with 8TX is lower than that with 4TX. This is because both 4TX and 8TX legacy codebooks have the same number of precoding matrices equal to 16. Therefore, 4TX benefits more from the legacy codebook than 8TX due to the more relative phase choices of each port. By contrast, the performance gain with SB proposed method (where $M_2 = M_3 \cdots = M_8 = 2$) improves much compared with WB legacy, i.e., around 2 dB gain. Thus we can infer that the 8TX BLER performance is more sensitive to the larger phase diversity compared to 4TX.

\section{Conclusion}
\label{sec:con}
In this paper, we investigated the codebook-based UL transmission performance for the single-layer using fully coherent antenna ports in the context of SB precoding. We analyzed the SB precoder selection criteria and designed an UL codebook used for SB precoding by increasing the number of relative phase shifts of each port. Via the simulations, we investigated the single-layer performance for different numbers of TX, including 8TX performance using the codebook standardized in the latest Release 18. We verified that the SB precoding method improves the performance gain compared with the current UL precoding scheme. Meanwhile, we illustrated that the performance gain is sensitive to the SB size selection as well as the relative phase shift diversity. 

In our future work, we will focus on the UL SB precoding performance using multiple layers. Besides, since the SB size selection matters for higher SB precoding efficiency, it is also worth designing an adaptive SB size scheme to deal with a varying channel whose spread delay is dynamic with time. 

\section{Acknowledgements}
\label{sec:ack}
The authors acknowledge the guidance of Sriharsha Ramarao Magani (MediaTek India Inc.) on the UL 8TX standardization and thank Meng-Che Chang (MediaTek USA Inc.) for insightful technical discussions throughout this project.

\bibliographystyle{IEEEtran}
\bibliography{ref}
\end{document}